\documentclass[pre, prb,reprint,superscriptaddress,twocolumn,amsmath,amssymb]{revtex4-1}

\usepackage{xcolor}
\usepackage{graphicx}
\usepackage{dcolumn}
\usepackage{bm}
\usepackage{bbm}
\usepackage{textcomp}
\usepackage[nice]{units}

\begin{document}
\renewcommand{\tensor}[1]{\underline{\underline{#1}}}
\renewcommand{\vec}[1]{\mathbf{#1}}
\title{Colloidal aggregates tested via nanoindentation and quasi-simultaneous 3D imaging}
\author{M. Roth}
\affiliation{Experimental Physics of Interfaces, Max Planck Institute for Polymer Research, Ackermannweg 10, 55128 Mainz, Germany}
\affiliation{Graduate School Materials Science in Mainz, Staudinger Weg 9, D-55128 Mainz, Germany}
\author{C. Schilde}
\affiliation{Institute for Particle Technology, TU Braunschweig, Volkmaroder Str. 5, 38104 Braunschweig, Germany}
\author{P. Lellig}
\affiliation{Experimental Physics of Interfaces, Max Planck Institute for Polymer Research, Ackermannweg 10, 55128 Mainz, Germany}
\author{A. Kwade}
\affiliation{Institute for Particle Technology, TU Braunschweig, Volkmaroder Str. 5, 38104 Braunschweig, Germany}
\author{G. K. Auernhammer}
\email[]{auhammer@mpip-mainz.mpg.de}
\homepage[]{http://www.mpip-mainz.mpg.de/~auhammer}
\affiliation{Experimental Physics of Interfaces, Max Planck Institute for Polymer Research, Ackermannweg 10, 55128 Mainz, Germany}
\date{\today}
\begin{abstract}
The mechanical properties of aggregated colloids depend on the mutual interplay of inter-particle potentials, contact forces, aggregate structure and material properties of the bare particles. Owing to this variety of influences experimental results from macroscopic mechanical testings were mostly compared to time-consuming, microscopic simulations rather than to analytical theories. The aim of the present paper was to relate both macroscopic and microscopic mechanical data with each other and simple analytical models. We investigated dense amorphous aggregates made from monodisperse poly-methyl methacrylate (PMMA) particles (diameter: $\unit[1.6]{\mu m}$) via nanoindentation in combination with confocal microscopy. The resulting macroscopic information were complemented by the three dimensional aggregate structure as well as the microscopic strain field. The strain field was in reasonable agreement with the predictions from analytical continuum theories. As a consequence the measured force-depth curves could be analyzed within a theoretical framework that has frequently been used for nanoindentation of atomic matter such as metals, ceramics and polymers. The extracted values for hardness and effective elastic modulus represented average values characteristic of the aggregate. On base of these parameters we discuss the influence of the strength of particle bonds by introducing polystyrene (PS) between the particles.
\end{abstract}
\pacs{}
\maketitle
\section{Introduction}
\label{sec:Introduction}
Suspensions of micro- and nanometer-sized particles (colloids) were intensively used to model atomic systems \cite{Science.301.5632,Science.296.5565,Science.440.319}. Increased lengthscales in case of colloids enable for optical observation with light microscopes and lead to slowed down kinetics compared to atomic systems. Capillary waves could be visualized in real space \cite{Science.304.847}, the dynamics of dislocation generation in weak colloidal crystals was studies on a single particle level \cite{Science.440.319} and experiments on colloidal glasses contributed to a better understanding of the glass transition \cite{PhysRevLett.99.028301,Science.287.5453}. These studies benefited from universal theoretical descriptions applied to both, colloidal and atomic systems. In particular mode coupling theory, initially developed for the glass transition in molecular liquids turned out to be suitable for the modeling of dense colloidal suspensions \cite{JPhysCondMatt.13.9113,JRheol.53.707}. In many of these systems the mechanical behavior is of central interest, as this characteristic is experimentally conveniently accessible. 

Besides this fundamental scientific motivation there is a technological interest in the mechanical properties of colloidal suspensions and in particular aggregated colloids. Colloids are frequently used as filler materials in various products of daily life. Depending on the application certain requirements related to morphology, abrasion resistance, specific surface and tendency to agglomeration must be met. These properties depend on the physio-chemical properties of the used materials as well as the technical control of the production process. Yet, in many cases the particles are not produced as single primary particles but rather as aggregates consisting of several primary particles. Additional mechanical treatments are required in order to obtain the primary particles or certain aggregate sizes. Generally, the influencing factors for breaking up the aggregates depend on their structural and material properties \cite{PhysRevE.79.061401} as well as the grinding and dispersing process \cite{ceramicforumIntenational.84.12,ChemEngSCi.65.3518}. Especially the micro-mechanical properties such as the breakage energy as well as plastic and elastic deformation energy at the nanoscopic scale have to be considered \cite{ChemEngTech.32.1521,ChemIngTech.81.1511}.

Ordered and disordered assemblies of colloids also attracted interest because of their specialized functionality and advantageous preparation in large scales and via self-assembly techniques. Examples are photonic colloidal crystals that show a great potential as optical elements \cite{ProcSPIE.7775.77750H,Nature.414.289} or surfaces that can be made super-hydrophobic by coating them with raspberry-like particles \cite{FaradayDiscuss.146.35}. Besides their primary function again the mechanical properties play a key role as these colloidal aggregates must render a certain mechanical stability for a convenient usage in daily life.

Such systems fundamentally differ from those mentioned in the beginning. Aggregated colloids are not in a dynamic equilibrium but are kinetically arrested. Bonds between the constituent particles last long and hinder structural rearrangements of the particles. Hence, any model based on the diffusive motion of the particles cannot be applied. Nevertheless, a comparison to equivalent atomic systems such as glasses, metals or ceramics provides other ways for a theoretical description. It is the aim of the present study to test the applicability of analytical continuum theory for the mechanical modeling of dense colloidal aggregates.

We chose nanoindentation as the mechanical test method. In a nanoindentation experiment a tip of defined geometry, spherical or triangular-pyramidal, is impressed into the specimen and retracted while the applied force is measured. Simple, yet quantitative analysis methods based on the continuum mechanical models of Boussinesq \cite{Boussinesq1885} and Hertz \cite{JReineAngewMath.156} were successfully applied to interpret nanoindentation data. Although the specimen was tested only very locally the extracted mechanical quantities like hardness and Young's modulus were in good agreement with those obtained from classical macroscopic testings \cite{ExpTech.33.66,JMatRes.1.601}. This might be related to the fact that despite of indentation depths down to $\unit[0.25]{\mu m}$ sufficiently large ensembles of atoms or molecules were tested to be treated as a continuum and obtain average material properties \cite{JMaterRes.21.3029}. 

In case of colloidal aggregates validity and applicability of these analysis methods had not been tested before and a priory could not be assumed. Even if the indentation depth was increased to several micrometers the number of involved particles was effectively decreased by a factor of roughly $10^{5}$ compared to atomic systems due to the size difference between atoms and colloids. Hence, the question arises whether the amount of tested material, i.e. the number of tested colloids, is large enough to properly define average material properties of the aggregate. The analogy is complicated even further if obvious differences in the interaction mechanisms between the constituent elements in atomic and colloidal systems are taken into account. Like in the dispersed state particles interact via electrostatic, steric and van der Waals forces. However, as soon as the particle aggregate structural properties become relevant, too. Dense and in particular crystalline structures show partial elasticity \cite{Nature.411.656} while fractal gels tend to be deformed plastically \cite{Langmuir.22.560}. If associated reorganization effects come into play material and surface properties of the particles need to be considered. Rough particles lead to increased frictional forces \cite{Langmuir.23.8392,PhysRevLett.83.3328,RevSciInstr.72.4164} that may favor rotational over sliding motions \cite{JApplPhys.104.054915}.

Although not directly detectable we accounted for these microscopic processes by complementing the bare force measurement with real-space imaging of the colloidal films using confocal microscopy. With the help of particle localization \cite{JCollInterfSci.179.298} and tracking algorithms \cite{PartTrack_Weeks} the particulate structure of the aggregate as well as three dimensional (3D) displacements of all individual particles were extracted during the whole indentation process. These displacements enabled for the calculation of the local strain tensor. Since this approach requires high 3D position resolution only few studies \cite{Science.318.1895,PhysRevE.81.011403} followed a similar approach. Yet, in these studies the results were successfully compared to continuum models, e.g. the nucleation dynamics of dislocation defects during the indentation of colloidal crystals \cite{NatMat.5.253,Science.440.319}. 

The paper is organized as follows. After a short repetition of the standard analysis procedure for nanoindentation data (section \ref{sec:TheoryOliverPharr}) we introduce the materials and experimental methods (section \ref{sec:Experimental}) used in this study. Then we will discuss the microscopic distribution of displacements and strains during the indentation process and compare it to theoretical predictions based on continuum mechanics (section \ref{sec:DisplStrain}). In sections \ref{sec:HE} and \ref{sec:Universality} the statistical analysis of the nanoindentation data will be presented and average material properties will be defined.
\section{Analysis of nanoindentation data}
\label{sec:TheoryOliverPharr}
\begin{figure}[tbp]
\includegraphics[width=0.38\textwidth]{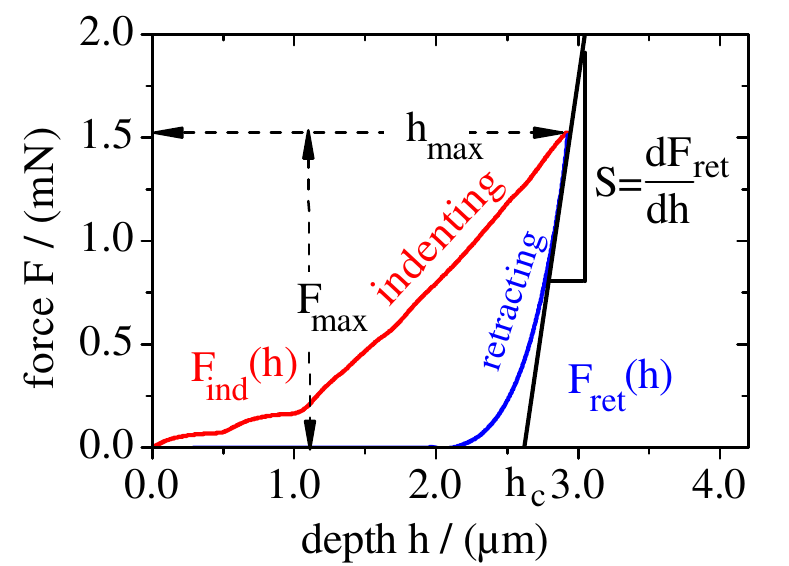}
\caption{Exemplary force-depth curve: See text for explanations of used variables.(After Oliver and Pharr \cite{JMaterRes.19.3}). 
}
\label{fig:TheoIndRealExa}
\end{figure}

A typical force-depth curve of a complete loading-unloading cycle in a nanoindentation experiment is shown in Fig.~\ref{fig:TheoIndRealExa}. Upon pressing into the sample the indenter deforms the sample both plastically and elastically. The applied force $F$ as well as the contact area between indentor and sample, projected to the surface along the indentation axis, $A$ increases steadily.  Both quantities are related with each other in terms of the hardness $H$ of the material.
\begin{eqnarray}
\displaystyle H&=&F/A
\label{eqn:hardness}
\end{eqnarray}
This relation for the hardness is mainly motivated from analogous macroscopic indentation tests after Vickers \cite{ProcInstMechEng.102.623} or Brinell \cite{Brinell.1901}
During the unloading process the elastic deformation is recovered resulting in a force pushing the indenter out of the specimen. This elastic recovery defines the reduced Young's Modulus $E_r$ via
\begin{eqnarray}
E_r&=&\frac{\sqrt{\pi}}{2}\,S/\sqrt{A}
\label{eqn:Youngred}
\end{eqnarray}
The contact stiffness $S$ is defined as the slope of the retraction curve $F_\text{ret}(h)$ at the maximum depth $h_\text{max}$:
\begin{eqnarray}
S &=& \left.dF_\text{ret}/dh\right|_{h=h_\text{max}}
\label{eqn:contstiff}
\end{eqnarray}
The Young's Modulus $E$ of the sample is obtained by eliminating the influence of elastic deformations of the tip ($E_i$, $\nu_i$):
\begin{eqnarray}
E&=&\dfrac{1-\nu^2}{1/E_r-(1-\nu_i^2)/E_i}
\label{eqn:Young}
\end{eqnarray}
Here $\nu$ and $\nu_i$ denote the Poisson ratios of sample and indenter, respectively. In contrast to the hardness the Young's modulus is well defined and equations \eqref{eqn:Youngred} to \eqref{eqn:contstiff} are based on the analytical continuum theories after Boussinesq \cite{Boussinesq1885} and Hertz \cite{JReineAngewMath.156}. Originally derived for the indentation of a spherical tip in an infinitesimally extended half space, it was shown that equations \eqref{eqn:Youngred} to \eqref{eqn:contstiff} also hold for other indentor geometries \cite{ZavodLab.41.1137,ZavodLab.53.76}. Moreover, Doerner and Nix proposed that $S$ is determined from a linear approximation to $F_\text{ret}$ at $h_\text{max}$ \cite{JMaterRes.1.601}. More recently, however, Oliver and Pharr pointed out that the steady reduction of the contact area $A$ while retracting the tip corrupts such a linear approximation \cite{Pharr}. Instead a power-law behavior
\begin{eqnarray}
F_\text{ret}&=&\alpha\cdot (h-h_f)^m
\label{eqn:PowerLaw}
\end{eqnarray}
is more accurate and the fitting parameters $\alpha$, $h_f$ and $m$ allow for a calculation of $S$.\\
Except for the projected contact area $A$ all quantities needed for the evaluation of hardness $H$ and Young's Modulus $E_r$ can be directly determined from the force-depth curve. In practice two routes are commonly applied to measure $A$, imaging of the residual indent via electron, light or surface probe microscopy or a mathematical procedure based on further assumptions about the elastic recovery. However, both approaches show major drawbacks and the method of choice depends strongly on the experimental details.

The mathematical determination of $A$ was proposed by Oliver and Pharr \cite{Pharr} and assumes a linear dependency of the restoring force on the indentation depth. Therefore, the effective depth $h_c$ of the tip in the sample is given by
\begin{eqnarray}
h_c &=& h_\text{max}-\epsilon\,F_\text{max}/S
\label{eqn:effdepthhc}
\end{eqnarray}
with a geometrical factor $\epsilon$. 
Once $h_c$ is known $A$ can be calculated from geometrical considerations \cite{ExpTech.33.66}, e.g. for a Berkovich indenter like in the present study:
\begin{eqnarray}
A &=& 24.56\,h_c^2
\label{eqn:contactarea}
\end{eqnarray}

Beside the normalized mechanical quantities $H$ and $E_\text{eff}$ the total, elastic and plastic deformation works can be used to characterize the mechanical properties of the specimen \cite{SurfCoatTech.204.2073,JMaterRes.17.502}. They are defined by the integrals of the indentation $F_\text{ind}(h)$ and retraction curves $F_\text{ret}(h)$ as follows:
\begin{eqnarray}
W_\text{tot} &=& \int_{0}^{h_\text{max}} F_\text{ind}(h) \,dh\\
W_\text{ela} &=& \int_{0}^{h_\text{max}} F_\text{ret}(h) \,dh\\
W_\text{pla} &=& W_\text{tot}-W_\text{ela}
\label{eqn:DefEnergy}
\end{eqnarray}

\section{Experimental}
\label{sec:Experimental}
	
\subsection{Sample Preparation}
\label{sec:ExpSample}
Colloidal aggregates were prepared as thin films. Compared to isolated aggregates such extended films are advantageous as the mechanical properties are not affected by finite size-effects and results from different indentation locations are well comparable.

The base particles were made from Poly(methyl methacrylate) (PMMA) and synthesized via dispersion polymerization with heptane as solvent \cite{CollPolySci.275.897}. A comb-like graft copolymer with a backbone of methyl methacrylate and glycidyl methacrylate and teeth made from poly(12-
hydroxystearic acid) \cite{CollSurf.17.67} was copolymerized with the PMMA and preferentially allocated at the surface of the forming particle. The poly(12-hydroxystearic acid) teeth served as a steric stabilization and prevented aggregation of the particles. Morevover, Nile red was added to the reaction mixture and homogeneously incorporated into the particles. The particles had a mean diameter of $\unit[1.6]{\mu m}$ and a polydispersity of $\unit[5]{\%}$ obtained from scanning electron microscope images on the basis of 100 particles. In order to minimize differences in the particle properties, all used particles originate from one single synthesis batch. 
 
For a first set of films these PMMA particles were dispersed in hexane (Sigma-Aldrich, $\unit[98]{\%}$ purity) at a low volume fraction of $\unit[2]{\%_\text{vol}}$. A $\unit[10]{\mu l}$ large droplet was placed and dried on a preheated glass substrate with $\unit[150]{\mu m}$ thickness \footnote{The thickness of the glass substrate is limited by the oil immersion objective used in the confocal microscope (section \ref{sec:ExpLSCM})} at an elevated temperature of $\unit[50]{^\circ C}$ without any further environmental control. Due to the fast evaporation and presumably strong convection during the drying process the resulting structure turned out to be completely amorphous.

\begin{figure}[tbp]
\includegraphics[width=0.38\textwidth]{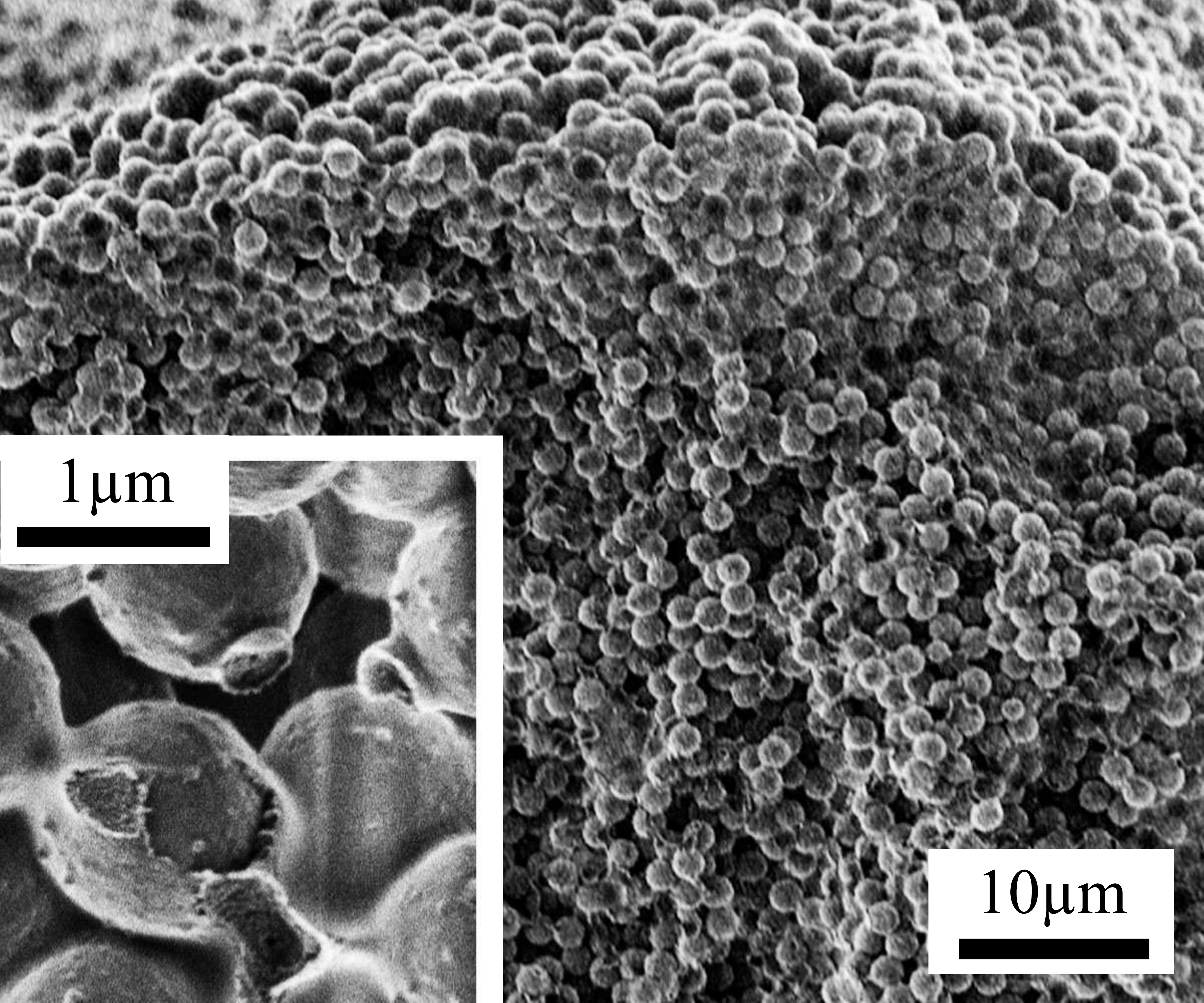}
\caption{Scanning electron microscope images of the surface and interior a fractured film made of PMMA particles with additional PS ($\unit[1]{wt\%}$ with respect to the amount of PMMA). The homogeneous distribution of PS throughout the whole film is visualized in the large image. The PS coverage of the PMMA particles and the solid state bridges are revealed if a particle is broken out of the film (see inset). }
\label{fig:SEMFilm}
\end{figure}

For a second set of samples a mixture of cyclohexyl bromide (CHB) and cis-decahydronaphtalin (DEC) was used to disperse the PMMA particles at a volume fraction of $\unit[10]{\%_\text{vol}}$. Various amounts of polystyrene (PS, $M_w=\unit[64]{kg/mol}, \,M_w/M_n=1.03$) were dissolved in the suspension to incorporate a tunable amount of solid bridges between the particles in the dried film. The mixing ratio of CHB and DEC was set to 4:1 by weight for practical reasons: Pure CHB swells and partially dissolves the particles while films casted from pure DEC were not stable and peeled off the glass substrate after evaporation. Moreover, the film surface was very smooth which might be related to the approximate density matching of the mixture and PMMA. In the dispersed state the dissolved PS acted as a depletion agent and induced attractive forces between the particles \cite{JChemPhys.113.10768}. As a result the particles aggregated to extended clusters \cite{PhysRevLett.96.028306}. Yet, these attractive forces were small compared to the hydrodynamic forces acting on the particles during drying at $\unit[50]{^\circ C}$. In the late stage of the drying process the remaining amount of CHB and DEC allocated on the surface of the particles
forming capillary bridges between them. The PS was strongly enriched in these regions forming solid glassy bridges (glass transition temperature of PS: $\unit[95]{^\circ C}$) when the solvent was completely evaporated. These bridges as well as the homogeneous average distribution of PS through the whole sample can be seen in the scanning electron microscope images in Fig. \ref{fig:SEMFilm}. 

\begin{figure}
\centering
\includegraphics[width=0.38\textwidth]{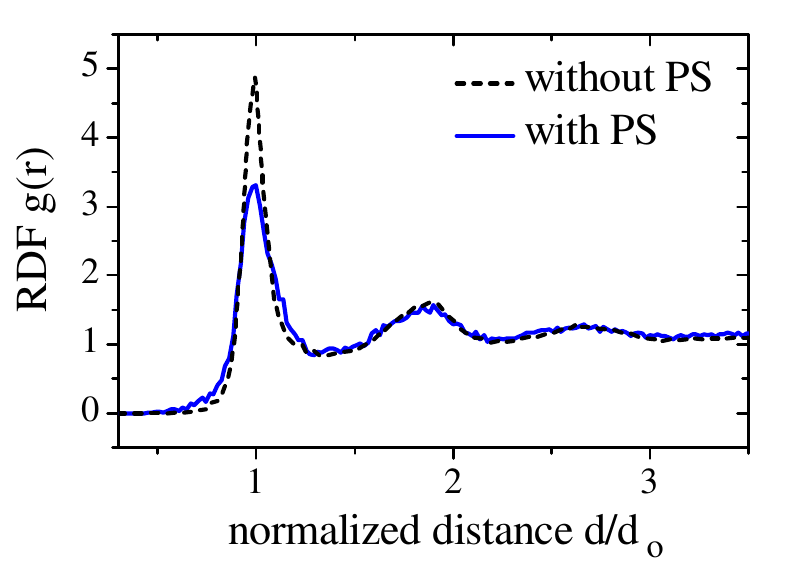}
\caption{Radial distribution functions (RDF) for the two types of amorphous colloidal films, without and with PS added to the PMMA particles. The distributions are rather similar showing the typical oscillation behavior. The normalization constant $d_o=\unit[1.6]{\mu m}$ was the same for both structures. Compared to the films with PS those without PS the first peak dropped in amplitude and slightly broadened. This indicates an overall smaller volume fraction of particles and local density variations, respectively \cite{PhysRevE.82.011403}.}
\label{fig:SamRDF}
\end{figure}

Further insight into the three-dimensional film structure was gained with the help of confocal microscopy (see section \ref{sec:ExpLSCM} for additional information). Infiltration of the film with a liquid that matches the refractive index of the PMMA significantly increased the image quality. We used so called laser liquids with a specified refractive index which are commercially available (refractive index: $n=1.5780\pm 0.0002$, Cargille Laboratories, USA). The films were neither re-dispersed nor were the solid state bridges affected which was checked up to several months after infiltration. The optical resolution was good enough to individually resolve the constituent particles. The thickness of the films ranged from $\unit[25]{\mu m}$ to $\unit[40]{\mu m}$ with local height variations of less than $\unit[2]{\mu m}$ over $\unit[10]{\mu m}$ in lateral dimensions. Using particle localization algorithms the 3D coordinates of all particles in the imaged volume were extracted. From these coordinates the radial distribution functions (RDF) \cite{EPL.84.46002} of both types of films were calculated (Fig.~\ref{fig:SamRDF}). 
Both distributions are very similar showing a strongly damped oscillation that is typical for an amorphous structure. However, a decrease and broadening of the first peak at $d/d_o=1$ was observed when introducing PS to the films. These variations are most likely caused by a decrease in the total volume fraction and by local variations in the particle volume fraction \cite{PhysRevE.82.011403}.


\begin{table}
\small
   \caption{\ Preparation methods and structural parameters of the colloidal films. The errors give a measure of the variations for different imaged regions of the sample.}
   \label{tab:PrepCompStruc}
   \begin{tabular*}{0.5\textwidth}{@{\extracolsep{\fill}}lllc}
   \hline
	Initial suspension	& Preparation					&Structure			&Particle				\\
						& method						&					&density 				\\
						&								&					&(\%vol)				\\
	\hline
	2\%wt PMMA			& fast drying					& amorphous			&$(63 \pm 2)$			\\
	in Hexane 			& at $\unit[50]{^\circ C}$		&					&						\\
	\hline
	10\%wt PMMA and		& fast	drying					&amorphous			&$(59\pm 3)$			\\
	0.02\%wt-0.1\%wt PS & at $\unit[50]{^\circ C}$		&     				&						\\
	in 4:1(vol) mixture &								&											\\
	of CHB and DEC		&								&					&		
   \end{tabular*}
\end{table}

The exact details on the particle interactions in the films were not investigated. The following assumptions were sufficient for the following discussion of the results. For the samples with added PS the particles predominantly interact with each other via solid bridges. These bridges have high elastic modules ($\sim\,GPa$) but are brittle and thus break upon small deformations. For the other samples dried from the suspension in hexane short-ranged contact forces of the small stabilization layer on the particle dominate the particle interaction. This interaction again defines a threshold value for bond rupture which yet is much smaller than that of the films with incorporated PS.

\subsection{Confocal Microscopy}
\label{sec:ExpLSCM}  
We used a home-made laser scanning confocal microscopes (LSCM) in fluorescence mode which is described in detail in reference \cite{arXiv:1106.3623v1}. The scan rate was one 2D frame per second with a sampling resolution of $\unit[0.1]{\mu m/pixel}$ in x and y-directions and $\unit[0.25]{\mu m/pixel}$ in z-direction. The microscope was equipped with an oil immersion objective (Olympus, UPlanApo PH3, NA=1.35) resulting in a spatial resolution of $\unit[250]{nm}$ in lateral directions (x, y) and $\unit[550]{nm}$ along with the optical path (z) if the refractive index was matched as mentioned above. Further general information about LSCM \cite{HandbookCM.Pawley,WEB.Minski.2008} as well as the special case of fluorescent labeled particles \cite{ApplOpt.24.4152,arXiv.1103.1051v1} can be found elsewhere.
  
The determination of the particle coordinates from the 3D microscope images was done using the algorithm by Crocker, Grier \cite{JCollInterfSci.179.298} and Weeks \cite{Science.287.5453}. In the first step the image quality was enhanced by a spatial filtering using a Gaussian kernel. Afterwards the particles appeared as isolated local intensity maxima which could be located subsequently. An additional position refinement step increased the precision of the particle coordinates to a level of better than $\unit[0.03]{\mu m}$. At this point the image analysis was finished and the colloidal film structure was well defined by a set of 3D particle coordinates.

\subsection{Nanoindentation}
\label{sec:Nanoindentation}
A simultaneous microscopic deformation analysis and force measurement was not possible with the present setup. Therefore, two complementary experiments were performed, force-sensitive nanoindentation with a commercial nanoindentor and so called ``live indentation'' in combination with confocal microscopy. 

\subsubsection*{Force-sensitive nanoindentation}
Force-sensitive nanoindentation experiments were done with a Tribo-Indenter\textsuperscript{\textregistered} TI 900 from Hysitron Inc. equipped with a Berkovich diamond tip. We chose a displacement controlled operation for which the tip was pressed into the sample at a constant indentation rate. Besides the depth of the tip inside the sample the normal force on the tip was recorded for the loading and unloading cycle resulting in so called force-depth curves. The maximal indentation depth and indentation velocity were kept constant for all samples at $\unit[3]{\mu m}$ and $\unit[0.5]{\mu m/s}$, respectively. In order to get an estimation of the standard deviation of the mechanical properties each sample was tested on a $5\times 5$ pattern of individual indentation spots that were separated by at least $\unit[30]{\mu m}$. After indentation the films were infiltrated with Cargille laser liquid and subsequently imaged with the confocal microscope in order to determine the extension of the residual indent. 

\subsubsection*{Live indentation}
For the live indentation experiments an indenter tip was mounted to a piezo-translation stage (Actuator PXY 200SG, Controller ENV40, piezosystems jena GmbH, Germany) which in turn was placed on the sample stage of the confocal microscope. The tip was pressed into the film in steps of $\unit[(0.5-1)]{\mu m}$. After each step the whole structure was imaged. During imaging the indenter was not moved. With the help of tracking algorithms \cite{PartTrack_Weeks} the trajectories of the particles could be followed through all measured 3D images. Thusly we obtained the vectorial displacement for each individual particle in the film for the complete indentation process. Besides a Berkovich tip also a spherical tip made from silica (diameter: $\unit[25]{\mu m}$) was used. The primary purpose of the live indentation experiments was a comparison of the deformation field with the analytical theory according to Hertz \cite{JReineAngewMath.156,ContactMechanicsChap4.Johnson} and Huber \cite{AnnPhys.316.153}. As this task was accomplished much easier for cylindrically symmetric indentation geometry we will focus on the results using the spherical tip. However, essentially the same observations were made for the Berkovich indenter as well. As the imaging and indentation occurs quasi-simultaneously the film needed to be infiltrated from the beginning.

\subsection{Microscopic deformation analysis}
\label{sec:DeformAnalysis}
\begin{figure}[tbp]
\includegraphics[width=0.38\textwidth]{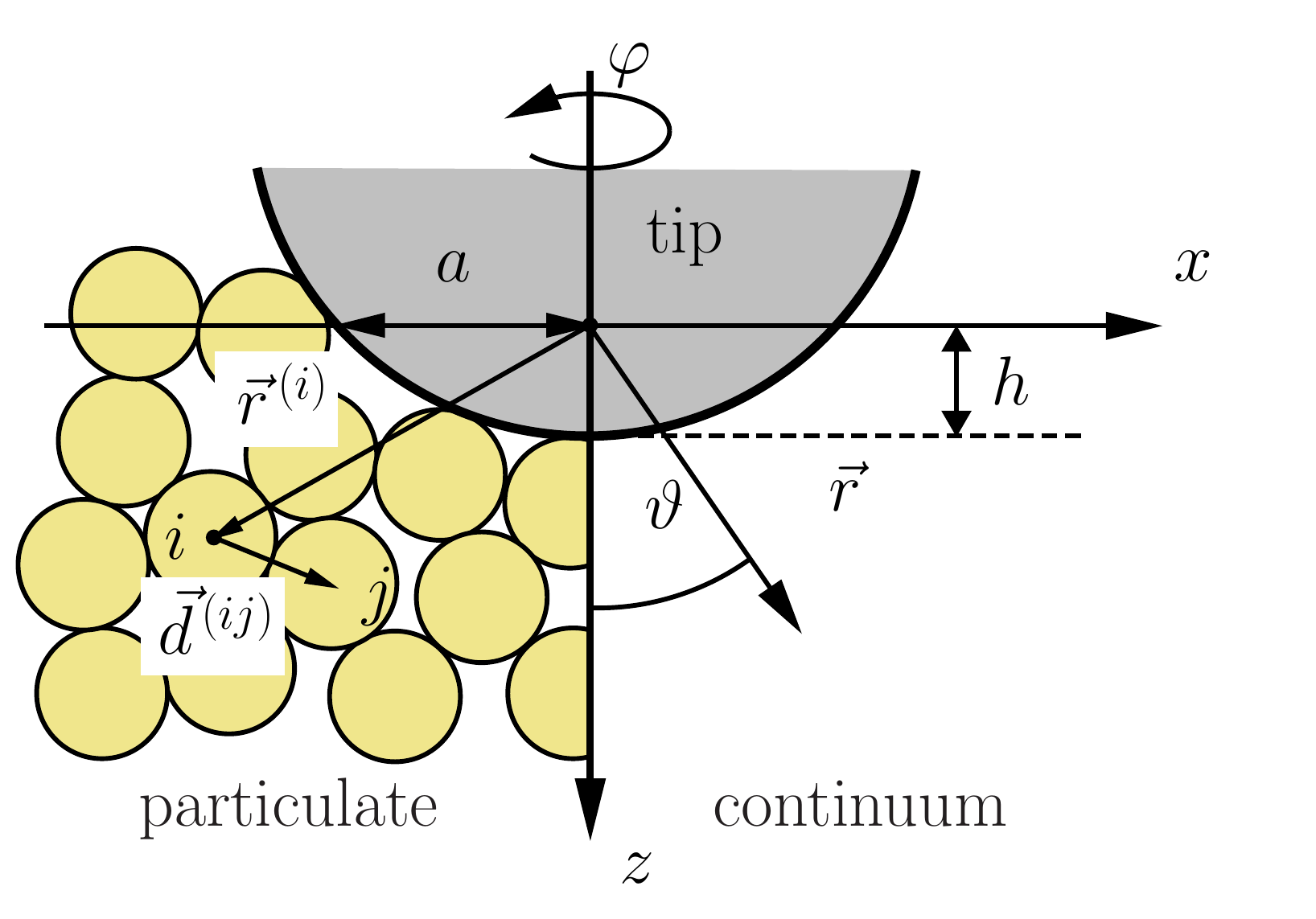}
\caption{Indentation geometry: The spherical indenter tip is indented in positive z-direction with $h$ denoting the depth of the tip inside the sample. Due to the cylindrical symmetry of the setup only the xz-plane is shown. We refer to the colloidal film either in the native particulate picture $\vec{r}^{\,(i)}$ or via continuous position vectors $\vec r$. Next to Cartesian coordinates also cylindrical coordinates $\vartheta,\,\varphi$ are introduced. }
\label{fig:DefIndentGeometry}
\end{figure}

In case of nanoindentation of atomic systems the applied external forces are transmitted through the specimen resulting in stress and deformation fields that are described as continuous fields. The relationship between these quantities is given by material properties like Young's and shear modules using the continuum elasticity theory, see e.g. Hertz \cite{JReineAngewMath.156,ContactMechanicsChap4.Johnson} and Huber \cite{AnnPhys.316.153}.\\
On the contrary in particulate systems stresses $\tensor{\sigma}^{(ij)}$ are localized to the contact points between two adjacent particles $i$ and $j$ (Fig.~\ref{fig:DefIndentGeometry}) and a priori need to be described discretely. 
Depending on the type and strength of the particle bonds these stresses result in a change of the bond vectors $\vec{d}^{\,(ij)}(h)$ relative to the bond vectors before indentation $\vec{d}^{\,(ij)}(\unit[0]{\mu m})$ 
\begin{eqnarray}
\vec{\Delta d}^{(ij)}(h) = \vec{d}^{\,(ij)}(h) - \vec{d}^{\,(ij)}(\unit[0]{\mu m})
\label{eqn:ChangeBonVec}
\end{eqnarray}
with $h$ denoting the depth of the indenter tip inside the film (Fig.~\ref{fig:DefIndentGeometry}). Yet, every particle is directly influenced by all of its $N_n$ nearest neighbors and particle rearrangements include rotations as well as tensile and compressive deformations of the local particle arrangement. We use the local elastic strain tensor $\tensor{\epsilon}^{(i)}$ \cite{PhysRevE.81.011403,Science.440.319} to quantify these deformations. It relates the changes in the bond vectors with the relative positions of the particles via an affine transformation. Any deviation from this affine transformation is caused by irreversible reorganizations of the particle structure and errors in the particle positions. So, $\tensor{\epsilon}^{(i)}$ can be determined by minimizing the expression
\begin{eqnarray}
\mathcal{F}^{(i)}(h) = \left|\sum\limits_{j=1}^{N_n} \Delta \vec{d}^{\,(ij)}(h) - \tensor{\epsilon}^{(i)}(h)\, \vec{d}^{\,(ij)}(\unit[0]{\mu m})\right|^2 
\label{eqn:straintensorpre}
\end{eqnarray}

In principle $N_n$ contains only particles, that are in physical contact with the central particle. However,  polydispersity of the particles complicates the decision which particles actually are in contact. Yet, we checked the results presented below for several values of the maximal distance between two neighboring particles and did not find a significant influence.

Although $\tensor{\epsilon}^{(i)}$ quantifies the deformation of the local particle structure it does not contain any information about the exact reorganization mechanisms. Elastic deformations of the particle bonds compete with more complex reorganizations like rolling and sliding if the particle bonds break. Moreover, deformations of the particles itself must be taken into account if the bonds are strong or local rearrangements are hindered.\\
We also calculated the vectorial displacement
\begin{eqnarray}
\vec{D}^{(i)}(h) = \vec{r}^{(i)}(h) - \vec{r}^{(i)}(\unit[0]{\mu m})
\label{eqn:CumDispl}
\end{eqnarray}
for particle $i$ and its absolute value $D^{(i)}=|\vec{D}^{(i)}|$. As $\vec{D}^{(i)}$ is based on relative particle positions, increased relative errors are expected compared to the pure particle coordinates. This is even more pronounced for $\tensor{\epsilon}$ that depends on changes in the relative positions. Therefore, both quantities were averaged over all particles that are located in a cubic subvolume with a side length of $\unit[3]{\mu m}$ which corresponds typically to an averaging over 8 particles. Moreover, we only discuss data at the maximal indentation depth of $h=\unit[3]{\mu m }$. In doing so we define quasi-continuous magnitudes $\vec{D}(\vec{r}\,)$ and $D(\vec{r}\,)$ as well as $\tensor{\epsilon}(\vec{r}\,)$ and $\mathcal{F}(\vec{r}\,)$ at the position $\vec{r}$. In the following we mostly show these averaged quantities in order to compare our findings with prediction of a continuum model of the indentation process.

\section{Results and discussion}
\label{sec:ResultsDisc}
\subsection{Displacement and strain field of the amorphous structure}
\label{sec:DisplStrain}
\begin{figure}[t]
\includegraphics[width=0.38\textwidth]{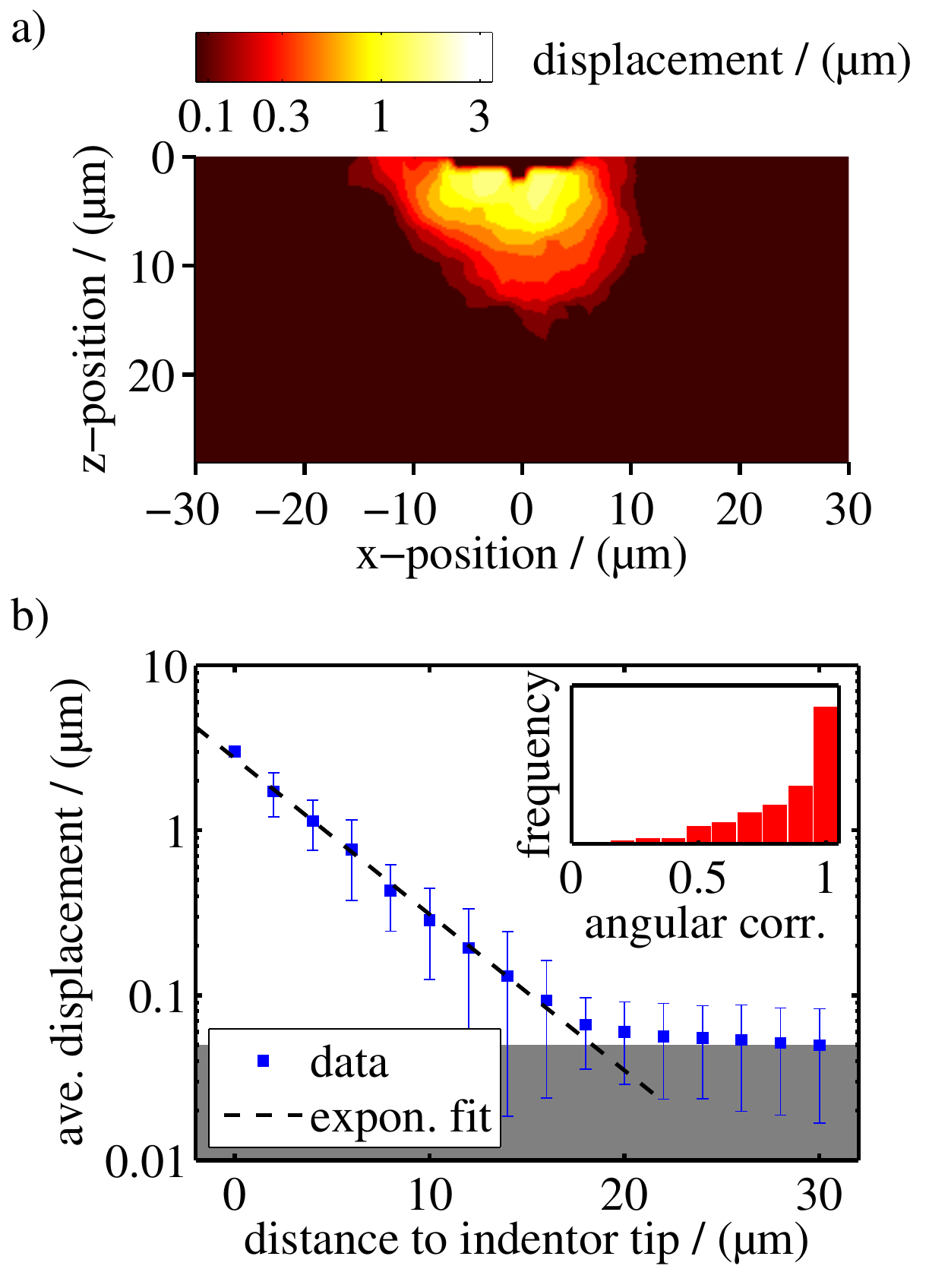}
\caption{a) Distribution of absolute particle displacements $D(\vec r\,)$ in the xz-plane in case of an amorphous structure at the deepest indent position of $\unit[3]{\mu m}$. b) The average absolute displacement shows an exponential dependency on the distance to the indenter tip. Due to the resolution limit of about $\unit[0.05]{\mu m}$ a determination of particle displacements is not possible for distances exceeding $\unit[20]{\mu m}$. The position vector $\vec{r}$ and the cumulative displacement vector $\vec{D}(\vec{r}\,)$ are mainly parallel as depicted in the inset in terms of the distribution of the projection $\vec{r}\cdot \vec{D}(\vec{r}\,) /(|\vec{r}|\,|\vec{D}(\vec{r}\,)|)$.}
\label{fig:IndentProfAmorph}
\end{figure}
We start the discussion of the results with the live indentation of the amorphous structure that was not reinforced by additional PS. As listed in table \ref{tab:PrepCompStruc} the volume density was close to the value of $\unit[63]{\%}$ expected for a random close-packed structure of monodisperse spheres. \\
Fig.~\ref{fig:IndentProfAmorph}a) shows the absolute cumulative displacement field $D(\vec{r})$ for a xz-cut through the film at an indentation depth of $\unit[3]{\mu m}$. The distribution was roughly cylindrically symmetric to the z-axis as expected for a spherical indenter tip and an amorphous structure. In this averaged representation no major spatial heterogeneities were apparent and the total displacement decayed monotonically with increasing distance to the indenter tip. Further averaging over all azimuth angles $\varphi\in[\unit[0]{^\circ},\unit[360]{^\circ}]$ and intervals of $\unit[15]{^\circ}$ in polar angles $\vartheta$ revealed an exponential dependency [see Fig.~\ref{fig:IndentProfAmorph}b)] with no significant dependency on $\vartheta$. The error bars represent the statistical standard deviation from the averaging process. Despite rather large values of the standard deviation in the order of the displacement itself the exponential fit is very good for the complete resolvable distance range. At a distance of about $\unit[20]{\mu m}$ the displacement dropped below the experimental resolution limit of the particle displacements of about $\unit[0.05]{\mu m}\approx\sqrt{2}\,\unit[0.03]{\mu m}$ corresponding to uncertainties in the particle positions mentioned above. This low value for the resolution limit suggests that the large standard deviations have a physical origin: they reflect a rather broad underlying distribution of displacements.\\
The overall uniformity of the displacement field was also observed for xy-cuts (not shown) at various depths. Displacement and particle position vectors were mainly parallel to each other. This is shown in the inset of Fig.~\ref{fig:IndentProfAmorph}b) in terms of the projection $\vec{r}\cdot \vec{D}(\vec{r}) /(|\vec{r}|\,|\vec{D}(\vec{r})|) $ defined as the normalized scalar product of both vectors. The histogram has a maximum at $1$ corresponding to fully parallel position and displacement vectors. 

\begin{figure}[t]
\includegraphics[width=0.38\textwidth]{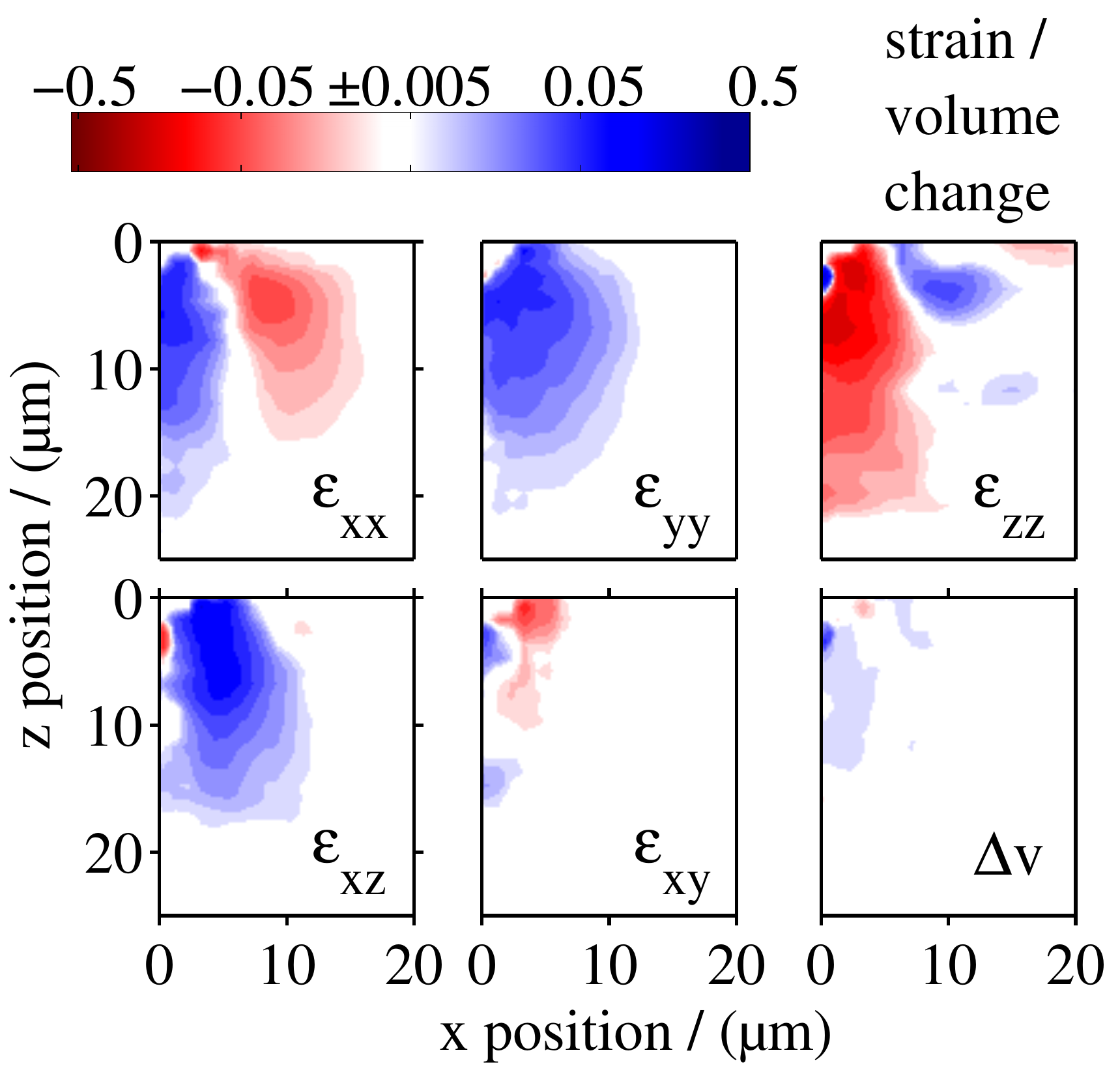}
\caption{Spatial distribution of the strain tensor elements $\epsilon_{\alpha\beta}$ at a depth of $\unit[3]{\mu m}$ for the case of a spherical tip indented in an amorphous film: The data is averaged for all possible values of $\varphi\in[\unit[0]{^\circ},\unit[360]{^\circ}]$ and therefore only shown for positive x-coordinates.}
\label{fig:IndentStrainTensorAmorph}
\includegraphics[width=0.38\textwidth]{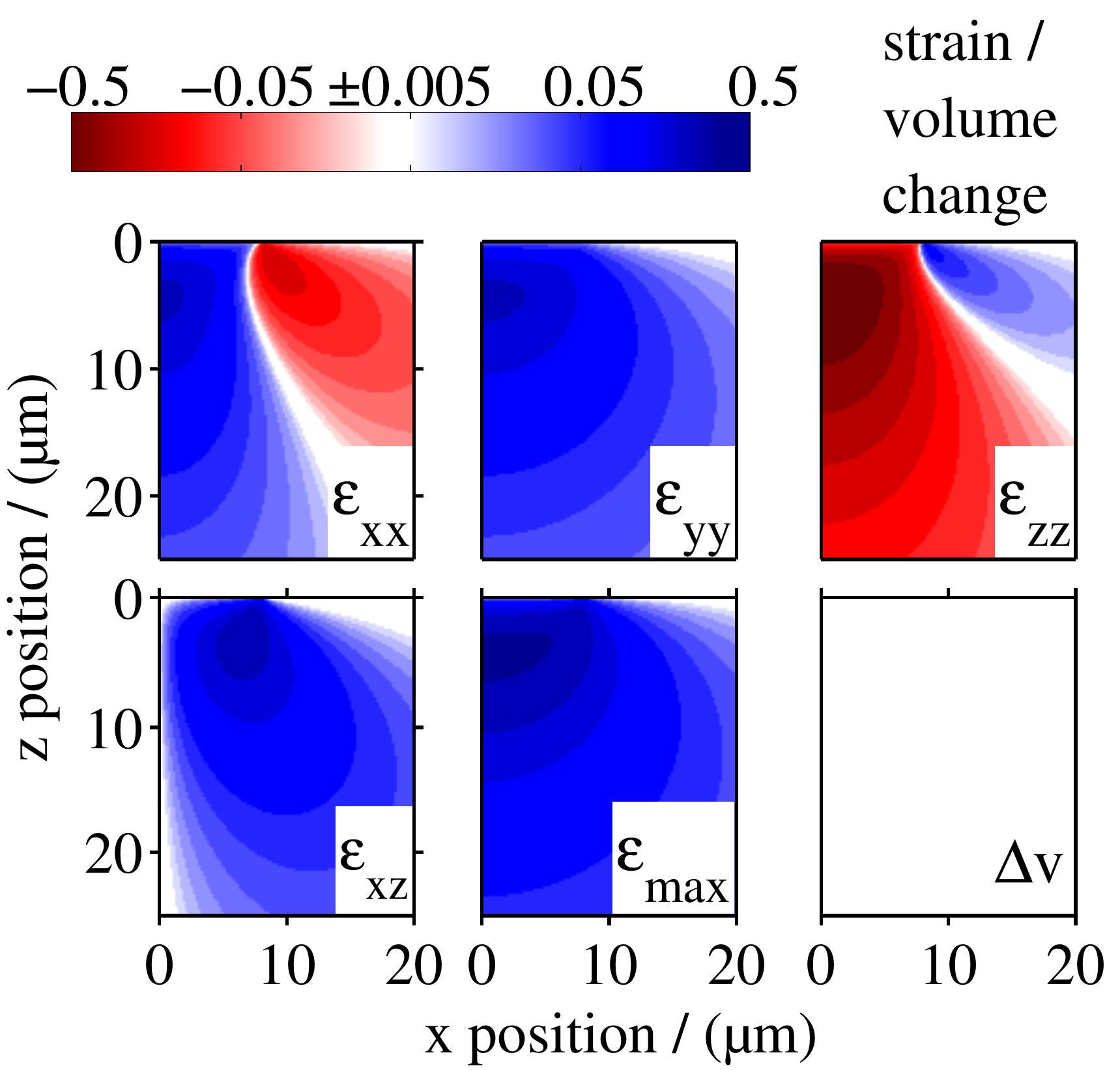}
\caption{Analytical calculation of the strain field based on an isotropic, elastic solid. Compare to Fig.~\ref{fig:IndentStrainTensorAmorph}. Since the component $\epsilon_{xy}$ vanishes, its graph is replaced by $\epsilon_\text{max}$. See text for further details.}
\label{fig:IndentStrainTensorAmorphCalc}
\end{figure}

More detailed information could be obtained from the strain tensor $\tensor{\epsilon}$. Because of azimuthal averaging only the positive x-axis is depicted in the graphs of $\tensor{\epsilon}$ at an indentation depth of $\unit[3]{\mu m}$ in Fig.~\ref{fig:IndentStrainTensorAmorph}. Just like the absolute displacement the absolute values for the strain components decreased with increasing distance to the indenter tip. Besides, there was a distinct patterning of positive and negative strains. In case of the diagonal elements 
$(\tensor{\epsilon})_{\alpha\alpha}\equiv\epsilon_{\alpha\alpha}$ positive and negative values represent dilation and compression of the local structure in direction of $\alpha$, respectively. Directly below the indenter the structure was strongly compressed in z-direction and dilated in x- and y-directions. The opposite was the case for polar angles $\vartheta \geq \unit[55]{^\circ}$ closer to the sample surface. Here the local particle arrangement was compressed in x-direction and dilated in y- and z-direction. Dilation and compression largely canceled each other leading to an almost vanishing relative volume change
\begin{eqnarray}
\Delta v = \epsilon_{xx}+\epsilon_{yy}+\epsilon_{zz}
\label{eqn:volchange}
\end{eqnarray}
Shear deformations also took place. The corresponding strain components in the xz-plane of Fig.~\ref{fig:IndentStrainTensorAmorph} were dominated by the xz-component. It was maximal at polar angles $\vartheta$ of about $\unit[55]{^\circ}$ for which the absolute values of $\epsilon_{xx}$ and $\epsilon_{zz}$ were minimal.

We compared the spatial distribution of the strain tensor elements with a continuum model by Hertz \cite{JReineAngewMath.156} and Huber \cite{ProcRSocLondA.197.416,AnnPhys.316.153} If the indented material was homogeneous and isotropic and the deformation was purely elastic, the stress tensor $\tensor{\sigma}$ could be calculated analytically and related to the strain tensor via Hooke's law
\begin{eqnarray}
\epsilon_{ij} = \frac{1}{E}\,\sigma_{ij} - \frac{\nu}{E}\,\left(\sum_k\sigma_{kk}\,\delta_{ij}-\sigma_{ij}\right)
\end{eqnarray} 
with $E$ and $\nu$ denoting the Young's modulus and Poisson ratio of the film.\\
The results of this calculation are shown in Fig.~\ref{fig:IndentStrainTensorAmorphCalc}. The overall agreement with the experimental data was best for a Poisson ratio of 0.5. This value represents an incompressible material that does not show any relative volume changes $\Delta v$ as it also was observed in the experiments. Moreover, the general agreement of calculations and experimental data in terms of the principle division into positive and negative values for the individual strain components $(\epsilon)_{ij}$ is clearly visible. Yet, the relative decay of the strain components with increasing distance from the indenter tip was much steeper for the experimental data. 

This discrepancy might be associated with the presence of the glass substrate which is not considered in the calculations. Finite size effects cannot be excluded but are probably less important as no influence of the actual sample thickness was observed experimentally. Another possible explanation is the presence of plastic deformations. As the bonds between the particles are rather weak for the films without PS, yielding of the film, i.e. irreversible particles rearrangement, must be considered. Yielding causes a release of the stress inside the film and thusly reduces the range of stress field compared to the calculations. Following Tresca's criterion for yielding such plastic deformations will first take place at the maximum of the principal stress difference \cite{ProcRSocLondA.197.416}. For a spherical indenter the maximum is located at a depth of $0.63\, a$ below the indenter tip with $a$ denoting the radius of the projected contact area (Fig.~\ref{fig:DefIndentGeometry}) \cite{ContactMechanicsChap4.Johnson}. Indeed such a localized yielding had already been found for other colloidal systems such as crystalline assemblies of soap bubbles \cite{Nature.411.656}. In our case the yielding point was close to the indenter tip even for the largest indentation depth ($0.63\,a \approx \unit[5]{\mu m}$) and masked by the limited spatial resolution after averaging. Moreover, if the yield stress was small, yielding could already take place at smaller indentation depths and therefore even closer to the tip where it could not be resolved. 

\begin{figure}[tbp]
\includegraphics[width=0.38\textwidth]{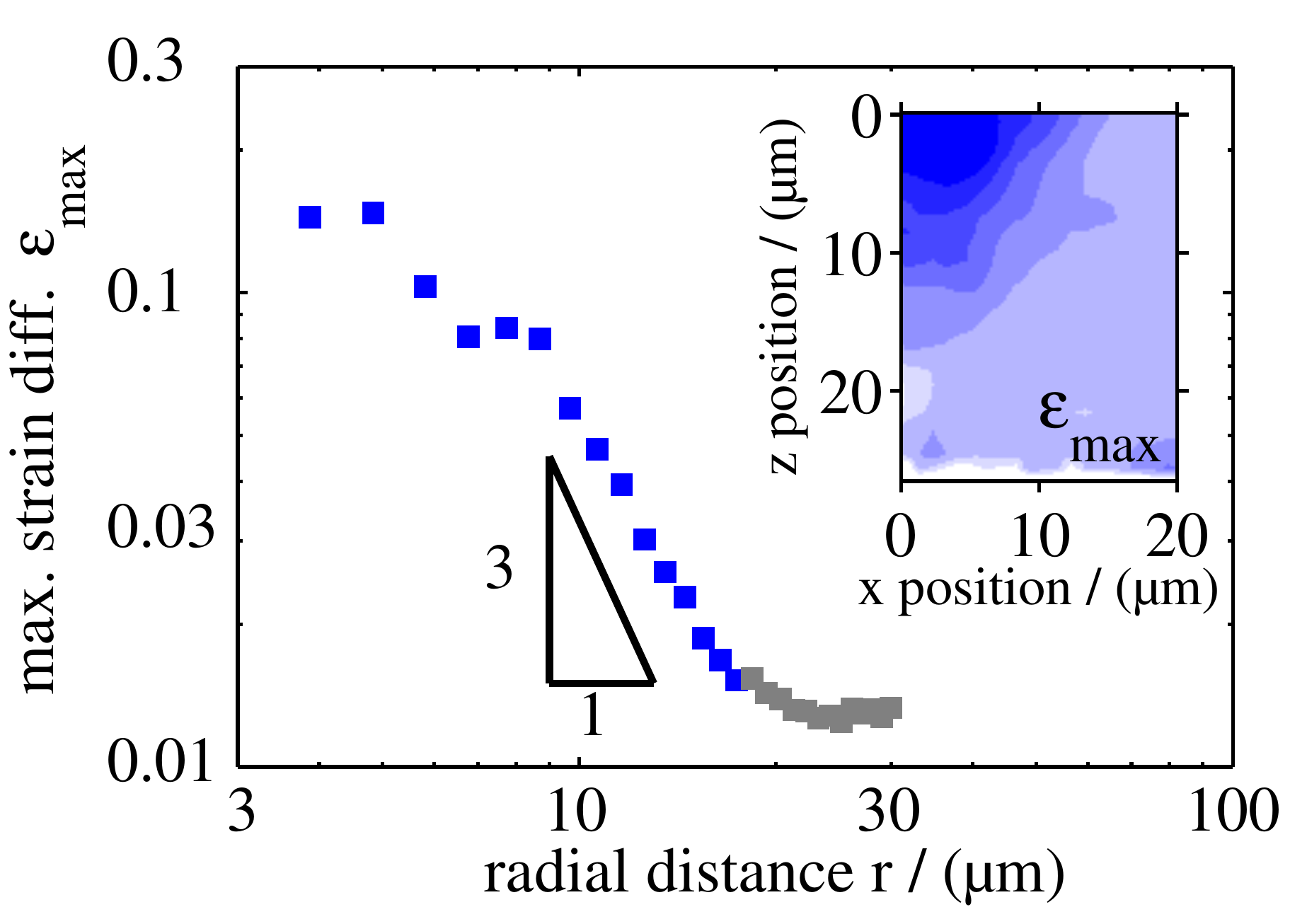}
\caption{The transition from a weak dependency of the maximal shear strain on the distance to the indenter tip to a power law behavior with a slope of approximately -3 is a hallmark of an elasto-plastic solid. The gray data points for the larger distances are masked by the resolution limit of the strain analysis. The complete xz-view is shown in the inset with a color-scale identical to Fig.~\ref{fig:IndentStrainTensorAmorph}}
\label{fig:StrainShowAmorphEpsMax}
\end{figure}

Hence, the colloidal film should rather be described as an elasto-plastic solid. In this model the strain in the specimen is divided into two regions \cite{ContactMechanicsChap6.Johnson,JMechPhysSolid.56.157}. At small distances from the indenter $r$ the local stress is large enough to exceed the yield stress of particle contacts. Plastic deformations take place by means of particle rearrangements and the rate of stress increase with further indentation decreases. At larger distances beyond the so called elastic-plastic boundary the local stress drops below the critical yield stress and the deformation is elastic. This region is characterized by a fast $r^{-3}$ dependency of the stress. Moreover, the distribution is merely independent of $\vartheta$. Experimentally this stress characteristics is best mapped by the maximal strain difference $\epsilon_\text{max}$ as it is a measure of the maximal stress difference \cite{ProcRSocLondA.197.416}. If the elements $\epsilon_{xy}$ and $\epsilon_{yz}$ are neglected with regard to $\epsilon_{xz}$\footnote{As the matrix representation of the strain tensor is symmetric it can be transformed to its principal axes. The non zero diagonal elements $\epsilon_{1},\,\epsilon_{2},\,\epsilon_{3}$ are connected with the maximal shear strain via $\epsilon_\text{max} = max(|\epsilon_{1}-\epsilon_{2}|,|\epsilon_{1}-\epsilon_{3}|,|\epsilon_{2}-\epsilon_{3}|)$. This definition leads to slightly larger values for $\epsilon_\text{max}$ but relative errors and the resolution limit are enlarged.}, $\epsilon_\text{max}$ is defined via
\begin{eqnarray}
\epsilon_{max}=\sqrt{(\epsilon_{xx}-\epsilon_{zz})^2/4+\epsilon_{xz}^2}
\label{eqn:MaximalShearStrain}
\end{eqnarray} 
The experimentally determined spatial distribution of $\epsilon_\text{max}$ as well as the distance profile (Fig.~\ref{fig:StrainShowAmorphEpsMax}) are in reasonable agreement with the described model. 

The comparison with the continuum model indicates that the measured strain tensor has elastic and plastic contributions. In the microscopic picture these contributions can be identified by rearrangements in the local particle structure. If the applied forces are small particle bonds are deformed elastically but when stressed beyond the yield point frictional sliding of particles occurs \cite{Langmuir.23.8392,PhysRevLett.83.3328,RevSciInstr.72.4164}. These processes are largely irreversible. Compressive strains might also have a contribution from deformations of the particles itself. This deformation $x$ is maximal for the particles that are in contact with the indenter tip. Under the assumption that a typical maximal load of $\unit[1.5]{mN}$ (Fig.~\ref{fig:TheoIndRealExa}) is distributed over roughly 25 particles, Hertzian contact theory \cite{JReineAngewMath.156,ContactMechanicsChap4.Johnson} predicts a particle deformation of:
\begin{eqnarray}
x = \left(\frac{9\,F^2}{16\,R^\star\,E^{\star\,2}_\text{part}}\right)^{1/3} \approx \unit[140]{nm} 
\end{eqnarray} 
Here $R^\star = (1/R_\text{part}+1/R_\text{ind})^{-1}$ denotes the effective particle radius defined by the radii $R_\text{part}$ and $R_\text{ind}$ of the particle and indenter, respectively. $E^\star_\text{part}\approx\unit[2.5]{GPa}$ is the effective Young's Modulus of the PMMA and the indenter tip material defined in analogy of equation \eqref{eqn:Youngred}. However, the corresponding contact pressure
\begin{eqnarray}
p=\frac{2}{\pi}\,E^\star_\text{part}\left(\frac{d}{R^\star}\right)^{1/2} \approx \unit[500]{MPa} 
\end{eqnarray} 
by far exceeds the compressive strength of PMMA ($\approx\unit[80]{MPa}$). So the particles are plastically deformed which manifests in a flattening of the particle surface at the contact point and the calculated deformation of $\unit[140]{nm}$ is overestimated. Hence, these particle deformations were present but only had a minor contribution to the strain close to the indenter tip in Fig.~\ref{fig:IndentStrainTensorAmorph}. As the force on the particles decayed steeply when going deeper in the film only the upper most particle layers were affected by plastic deformations. The lower particles showed predominantly elastic deformations that could not be distinguished from structural reorganizations from the confocal images alone. These considerations are also in accordance with the observation from above, that the film is incompressible.
\subsection{Hardness and effective Young's modulus}
\label{sec:HE}
The possibility to model the microscopic deformation of a the film with continuum mechanical models suggest, that the force-depth curves can be analyzed in the framework described in section \ref{sec:TheoryOliverPharr}. However, at a closer look a few difficulties arise. 
 
At first the force-depth curves need to be corrected for the deformation of the thin glass substrate. Upon indentation the substrate deforms just like the sample and thusly reduces the actual indentation depth. The higher the indentation force the stronger the reduction is. Indentation on a bare glass substrate showed that the deformation is highly elastic and an effective spring constant of $\unit[20]{N/mm}$ was obtained from the unloading part of the indentation process. The deformation of the substrate was calculated from the applied force according to Hook's law and subtracted from the raw indentation depth.

Further problems arise from the imprecise starting point of the force-depth curves. Upon approaching to the sample surface the tip might indent directly on top of a particle or into the interstice between them. After the initial contact with the tip the upper most particles rearranged to fit the shape of the indenter. As only few particles were involved this first consolidation process had a diffuse mechanical response that differed strongly from one indentation spot to the other. Large strains in the vicinity of the indenter tip were revealed in live indentation experiments and eventually led to the two small humps in Fig.~\ref{fig:TheoIndRealExa} in the early stage of the indentation. Given the fact that the particle radius of $\unit[0.8]{\mu m}$ was not far off the total indentation depth $h_\text{max}$ these onset uncertainties gave rise to large errors in $h_c$ and even more pronounced in $A$ according to equations \eqref{eqn:effdepthhc} and \eqref{eqn:contactarea}. The problem of an unclear total indentation depth was even more intensified when adhesion came into play and single particles were pulled out of the film and adhered to the indenter tip. 
These adhered particles could cause a premature start of the subsequent indentation process and with this again gave another source of errors in $h_\text{max}$.

\begin{figure}[tbp]
\centering
\includegraphics[width=0.25\textwidth]{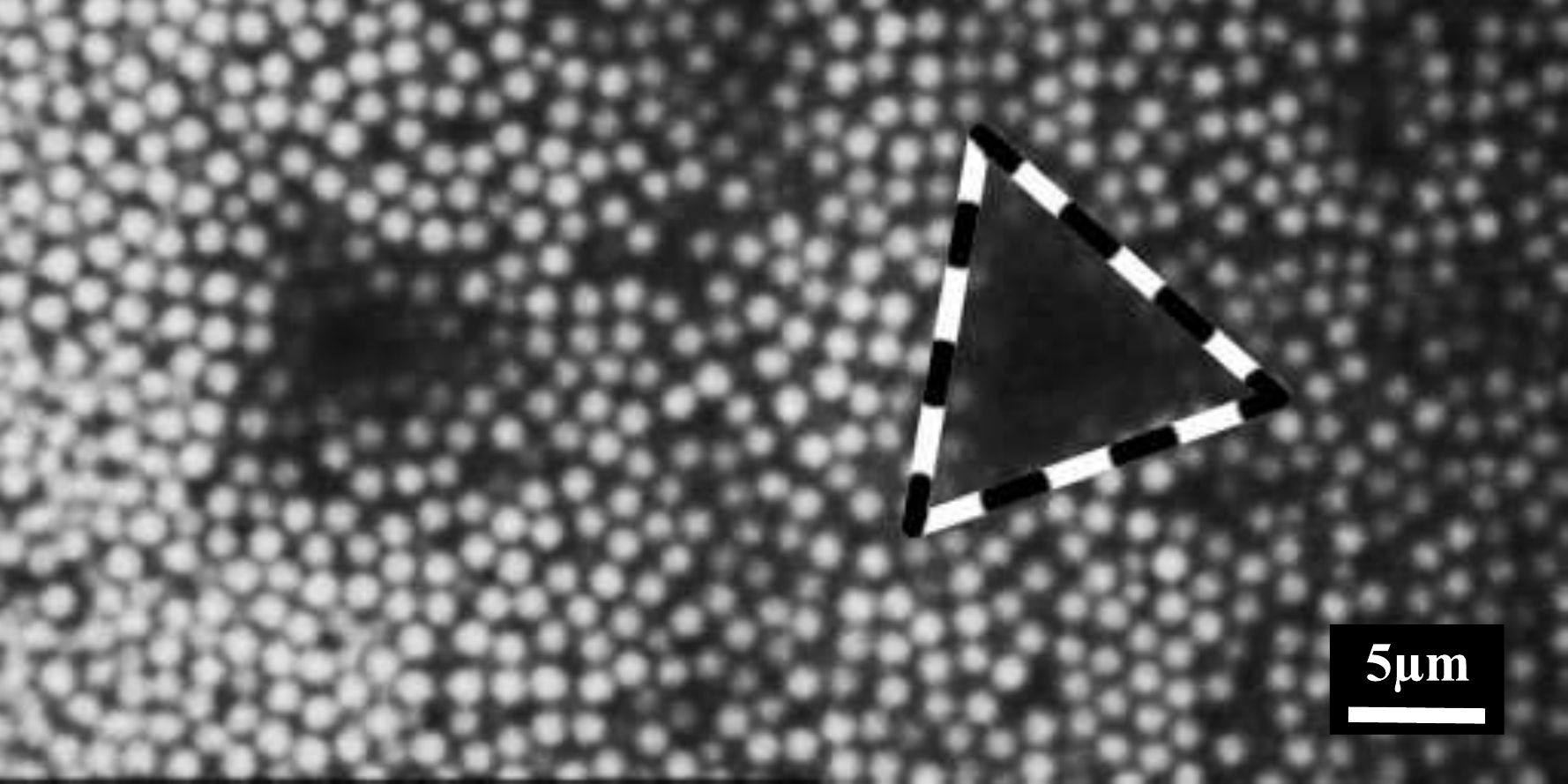}
\caption{The confocal image shows the colloidal film after indentation. The two neighboring indent appears as an dark area devoid of bright particles. The triangle indicates the shape of the indenter tip.}
\label{fig:Indentxy}
\end{figure}
 
We bypassed these difficulties by choosing the second possibility to determine the contact area $A$ by imaging the remnant indent. However, this method disregards the elastic recovery of the material upon unloading. Yet, the live indentation experiments showed that only particles with small displacements at larger distances to the tip return to their initial locations. On the contrary particles in direct contact with the indenter were largely irreversibly displaced and mapped the tip geometry. Fig.~\ref{fig:Indentxy} shows a xy slice of a complete 3D confocal data set showing two neighboring indent positions. The triangular shape of the Berkovich indenter is clearly visible and a determination of $A$ is feasible.  

\begin{figure}[tbp]
\includegraphics[width=0.38\textwidth]{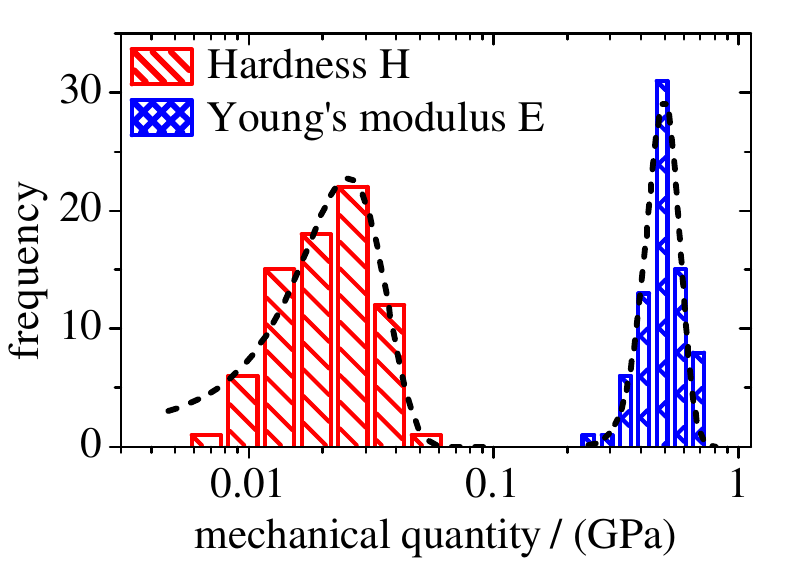}
\caption{Histograms}
\label{fig:AmorphHist}
\end{figure}

\begin{figure}[tbp]
\includegraphics[width=0.38\textwidth]{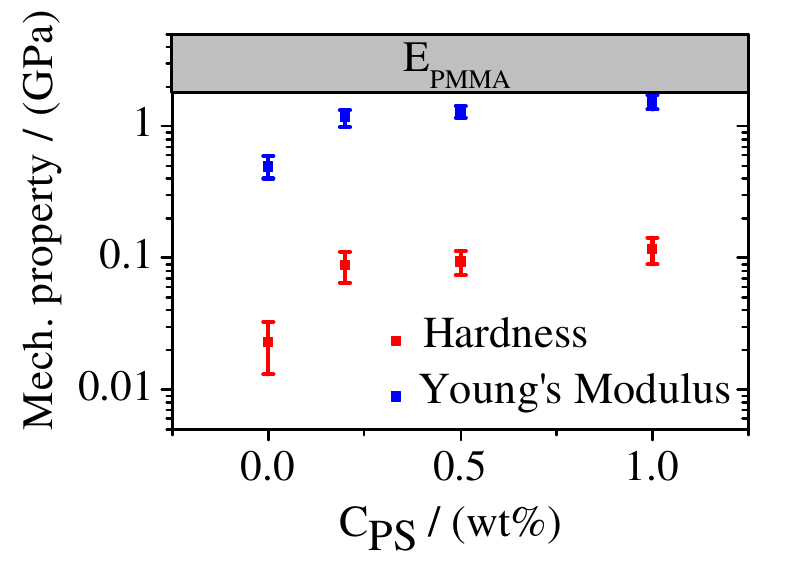}
\caption{Hardness and Young's modulus show the same behavior for a variation of binding strength between the particles in the film. Upon a sharp increase for the smallest amount of added PS, both similarly approach a plateau.}
\label{fig:HardYoungPoly}
\end{figure}

The histograms of the extracted values for hardness $H$ and effective modulus $E_\text{eff}$ of the amorphous films without additional PS are shown in Fig.~\ref{fig:AmorphHist}. Although the distributions are rather broad, they can be well described by a single Gaussian function each. The mean values are given by $\overline{H}=\unit[0.025
\pm 0.010]{GPa}$ and $\overline{E}=\unit[0.50 \pm 0.10]{GPa}$.  The same is true for the films with additional PS: the distributions are monomodal. However, already the smallest amount of PS led to a sharp increase in hardness and elastic modules by factors of 5 and 2.5, respectively (Fig.\ref{fig:HardYoungPoly}). Remarkably, additional PS did not cause any further enforcement of the film. 

Due to the lack of other approved mechanical test methods we can compare the observed mean values only with the corresponding literature values for bulk PMMA ($H_\text{lit}=\unit[1.9]{GPa}$ \cite{JApplPhys.90.1745} and $E_\text{lit}=\unit[(1.8-3.1)]{GPa}$). The hardness of a material is defined as the force per contact area that is needed to deform the specimen irrespective of the type of deformation. Irreversible rearrangements of particles are not distinguished from elastic deformation of particles and bonds. In colloidal films without additional PS rearrangements of the particles are promoted by a small bond strength leading to a relaxation of stresses. As a result the force required to indent the film is largely reduced compared to bulk PMMA. The hardness is decreased by almost two orders of magnitude compared to the literature values. The situation is different in case of the films with PS where strong solid bridges between the particles hinder rearrangements. We stress on the fact that the interaction between the particles is still mainly based on van der Waals forces. Hence, the strength of the particle bond is a mere effect of the increase contact area between adjacent particles mediated by the PS. As a result the hardness is increased but still far below the value for bulk PMMA.

The Young's modulus $E$ is a measure of elastic recovery in the very beginning of the retraction cycle. In this sense it is directly correlated to the elastic energy that is stored in the deformation of the contact bonds and the particles itself. If the particles do not rearrange or deform plastically during the indentation a larger amount of elastic energy can be accumulated and recovered when the tip is retracted from the sample. As these rearrangements are partially hindered for the films with additional PS due to the increased bond strength, $E$ is larger than for those without PS. However, further addition of PS does not affect the mechanical properties of the enforced films. This behavior can be understood, if one assumes that $E$ of the films with PS is effectively determined by the elastic properties of PMMA and PS themselves. As the Young's modules of PS and PMMA at room temperature are in the same order of magnitude $E$ of the film depends mainly on the total volume fraction of PMMA and PS together. This quantity in turn is supposed to be rather independent of the relative amount of PS since PS and PMMA distribute homogeneously during the preparation of the films. This model is further supported be the fact, that $E$ for the films with PS is very close to the literature value of bulk PMMA.

\subsection{Universality}
\label{sec:Universality}
\begin{figure}[tbp]
\includegraphics[width=0.38\textwidth]{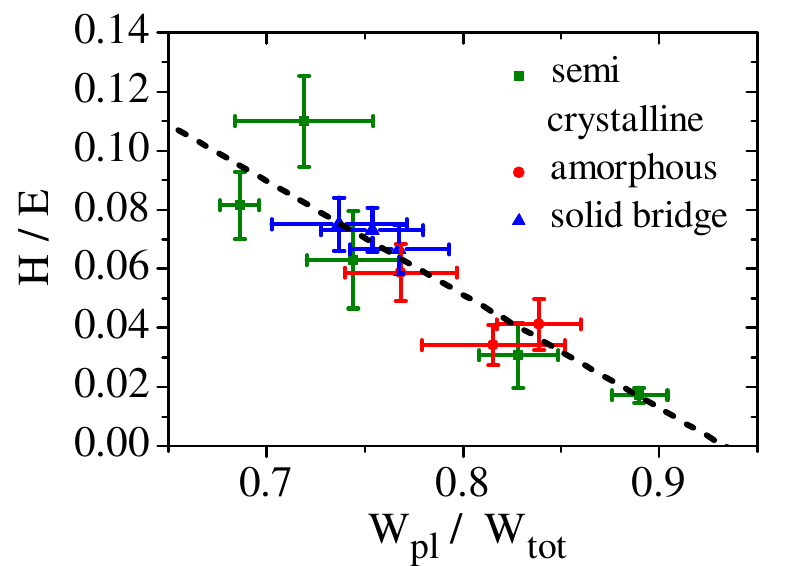}
\caption{Irrespective of structure or binding strength the mechanical properties of all samples can be universally related with each other in a plot of the ratio $H/E$ versus the fraction of plastic deformation work $\mathcal{W}_\text{pla}\,/\,\mathcal{W}_\text{tot}$.}
\label{fig:Universal}
\end{figure}
Further checks for the consistency of the used analysis procedures were found in the previous works \cite{ApplPhysLett.73.614} that related the mechanical quantities $H$ and $E$ to the elastic and plastic fractions of the deformation work. Scaling arguments predicted a linear correlation of $H/E$ and $\mathcal{W}_\text{pla}/\mathcal{W}_\text{tot}$ irrespective of the used materials \cite{SurfCoatTech.135.60,SurfCoatTech.204.2073}. Via this relation metals, ceramics and glasses could be universally superimposed on a single curve. The same was true for our colloidal films. Films with and without PS collapsed to a single line with a slope of $(-0.35\pm 0.03)$ (Fig.~\ref{fig:Universal}). Even semi-crystalline colloidal films, which will be discussed in detail in further publications, follow the same relation. The values for the slope differs substantially from the corresponding value reported in literature of $0.2$ \cite{JMaterRes.19.3} for atomic systems. This, however, might be caused by systematic errors in the values for the hardness and Young's modulus as possible sources for these uncertainties were discussed above. Only further extensive experiments or simulations can uncover the exact origin of the deviation.

\section{Conclusions}
\label{sec:Conclusions}
We presented a study of the mechanical properties of colloidal aggregates tested via nanoindentation. Although the sample was tested only very locally the induced deformation was in qualitative agreement with the predictions of a continuum theory as well as with indentation of atomic matter. As a result the theory of Oliver and Pharr was successfully applied to obtain average material properties. Unfortunately, there is a lack of other methods to verify the absolute values. Hardness and effective elastic modulus showed a strong dependency on the strength of the particle bonds as it has strong impact on the dominant deformation processes during the indentation. 

In general nanoindentation proved to be a potent method for mechanical characterization of colloidal films. Besides a simple experimental procedure the measurement data can be processed with standard analysis methods. Despite these benefits and the combination with quasi-simultaneous confocal microscopy no information about the microscopic forces between the particles and the exact reorganization processes could be gained. For this reason also the origin of the large statistical scatter in the average displacements of Fig.~\ref{fig:IndentProfAmorph}\,b) or the mechanical quantities in Fig.~\ref{fig:AmorphHist} remains unclear. Whether a broad distribution of contact forces like in force chains can account for it needs to be clarified.

We will complement our findings with discrete element method (DEM) calculations and further experimental studies in future publication. The DEM calculations can be used to simulate the microscopic deformation field as well as force-depth curves on the base of the mechanical properties of the constituent materials and thus is complementary to the presented experiments. On the other hand, in future experiments we will use hollow spheres as local force sensors \cite{Langmuir.25.2711} and anisotropic labeled particles for sensing individual particle rotations \cite{arXiv:1106.3623v1} in order to get a deeper insight into the reorganization processes. 
\section{Acknowledgements}
We appreciate fruitful discussions with Doris Vollmer and Thomas Palberg. Financial support by the Deutsche For\-schungs\-gemein\-schaft through the SPP 1486 and the SFB TR6 is gratefully acknowledged by G.K.A. and P.L., respectively. M.R. is a recipient of a fellowship through funding of the Excellence Initiative (DFG/GSC 266).  

\bibliography{literature_vs03}

\end{document}